\documentclass[prb,twocolumn,amsmath,superscriptaddress]{revtex4}
\usepackage{epsf,amssymb}

\usepackage{bm}

\begin{document}

\title{Fine structure of excited excitonic states in quantum
disks}

\author{M.~M.~Glazov}

\author{E.~L.~Ivchenko}

\affiliation{A.F. Ioffe Physico-Technical Institute RAS, 194021
St.-Petersburg, Russia}

\author{R.~v.~Baltz}
\author{E.~G.~Tsitsishvili}

\affiliation{Universit\"{a}t Karlsruhe, 76128 Karlsruhe, Germany}

\begin{abstract}
{We report on a theoretical study of the fine structure of excited
excitonic levels in semiconductor quantum disks. A particular
attention is paid to the effect of electron-hole {\it long-range}
exchange interaction. We demonstrate that, even in the
axisymmetric quantum disks, the exciton $P$-shell is split into
three sublevels. The analytical results are obtained in the
limiting cases of strong and weak confinement. A possibility of
exciton spin relaxation due to the resonant LO-phonon-assisted
coupling between the $P$ and $S$ shells is discussed. }
\end{abstract}

\maketitle

\section{Introduction}
In the envelope-function approximation, the excitonic states in
semiconductors and semiconductor quantum dots can be classified by
referring to the orbital shape of the exciton envelope function
($S$-, $P$-, $D$-like) and the character of the electron and hole
Bloch functions (bright and dark states). For example, the state
$| P_x y \rangle$ means a bright $P_x$-shell exciton with the
electron-hole dipole moment directed along the $y$ axis.
Theoretically, the fine structure of excited states of
zero-dimensional (0D) excitons has been studied for excitons
localized, respectively, by rectangular islands in a quantum-well
structure \cite{exchjetp}, quantum disks with a Gaussian lateral
potential \cite{taka} and lens-shaped quantum dots \cite{zunger}.
However, up to now the splitting of excited-exciton levels has
been analyzed for a fixed orbital shell, e.g., the $P_x$ shell,
which is valid in the case of strongly anisotropic confinement so
that the orbital splitting of the $P$-like shells $P_x$ and $P_y$
exceeds by far the exchange-interaction energy. Here we show that,
for axially symmetric or square quantum dots, one has, in
addition, to take into account the exchange-interaction-induced
mixing between the excitonic states $| P_x y \rangle$ and $| P_y x
\rangle$. Moreover, for the first time we focus on the interplay
between the exchange interaction and anisotropic shape of the dot
and develop an analytic theory in the particular case where the
orbital and exchange splittings are comparable.
\section{Electron-hole exchange interaction in quantum disks}
The two-particle excitonic wave function can be written as a
linear combination of products $\Psi_{sj}({\bm r}_e, {\bm r}_h) |
s , j \rangle $, where $| s , j \rangle$ is a product of the
electron and hole Bloch functions, $s$ and $j$ are the electron
and hole spin indices, $\Psi({\bm r}_e, {\bm r}_h)$ is the
envelope function, and $\bm r_{e,h}$ is the electron (hole) 3D
radius-vector. In the following, for the sake of simplicity, we
concentrate on heavy-hole excitons with $j = \pm 3/2$ and the
long-range mechanism of electron-hole exchange interaction which
allows to discuss only the bright excitonic states $| s , j
\rangle$ with $s + j = \pm 1$ or their linear combinations
$|\alpha\rangle$ with the dipole moment $\alpha=x,y$. The
short-range mechanism may be taken into consideration similarly to
\cite{kaminsk} if one includes an admixture of light-hole states
into the heavy-hole exciton wave function.

We consider a quantum disk formed by a 2D harmonic potential
$V(\bm \rho_e,\bm \rho_h) = A_e\rho_e^2 + A_h\rho_h^2$ in a
quantum well grown along the $z$-axis. Here the 2D vector $\bm
\rho_{e,h}$ determines the in-plane position of an electron or a
hole, and $A_{e,h}$ are positive constants. Assuming that the
confinement along the growth direction is stronger than both the
quantum-disk and Coulomb potentials the envelope for the
electron-hole pair wavefunction can be written as
\begin{equation}\label{psieh}
\Psi(\bm r_e,\bm r_h) = \psi(\bm \rho_e,\bm \rho_h)
\varphi_e(z_e)\varphi_e(z_h),
\end{equation}
where $\varphi_{e,h}(z_{e,h})$ are the respective $z$-envelopes of
electron and hole, and $\psi(\bm \rho_e,\bm \rho_h)$ is an
in-plane wavefunction of exciton calculated with allowance for
both Coulomb interaction and quantum disk potential. The form of
the function $\psi(\bm \rho_e,\bm \rho_h)$ depends on the
relationship between the potential radius and 2D-exciton Bohr
radius, $a_{\rm B}$. In the {\it weak} confinement regime, i.e. in
a quantum disk with the diameter exceeding $a_{\rm B}$, the
two-particle envelope probe function $ \psi(\bm \rho_e,\bm
\rho_h)$ is a product $F(\bm R) f({\bm \rho})$ of two functions
describing, respectively, the in-plane motion of the exciton
center of mass $\bm R=(X,Y)$ and the relative electron-hole
motion, ${\bm \rho} = {\bm \rho}_e - {\bm \rho}_h$. On the other
hand, in small quantum disks with {\it strong} confinement where
single-particle lateral confinement dominates over the Coulomb
interaction, the pair envelope can also be presented as a product
of two functions, $\psi_e({\bm \rho}_e) \psi_h({\bm \rho}_h)$, but
here they describe the independent in-plane localization of an
electron and a hole.

For the exciton envelope functions presented in the form
(\ref{psieh}), the matrix element $\mathcal H^{(\rm long)}_{n'n}$
of long-range exchange interaction taken between the exciton
states $n'$ and $n$ is written as follows
\begin{equation}\label{work}
\frac{1}{2\pi\varkappa_\infty} \left(\frac{e\hbar |p_0|}{m_0 E_g}
\right)^2 \int d\bm K \frac{K_\alpha K_{\alpha'}}{K}
\widetilde{\psi}^{*}_{n'}(\bm K) \widetilde{\psi}_n(\bm K),
\end{equation}
where the exciton-state index $n$ includes the dipole moment
$\alpha$, $\varkappa_\infty$ is the high-frequency dielectric
constant, $m_0$ is the free electron mass, $E_g$ is the band gap,
$p_0$ is the interband matrix element of the momentum operator,
and we introduced the 2D Fourier-transform
\[
\widetilde{\psi}(\bm K) =   \int d\bm R\: e^{-i\bm K \bm
R}\psi(\bm R,\bm R)
\]
of the function $\psi(\bm \rho_e,\bm \rho_h)$ at the coinciding
coordinates, $\bm \rho_e = \bm \rho_h \equiv {\bm R}$.

\section{$P$-Orbital excitons in axially-symmetric disks}
We start with an axially-symmetric quantum disk and calculate the
fine structure of the $P$-orbital exciton level and then proceed
to a slightly anisotropic disk. The straightforward calculation
shows that the long-range exchange Hamiltonian (\ref{work}) has
the following non-zero matrix elements
\begin{equation}\label{psplitting}
\langle P_x x|\mathcal H^{(\rm long)} | P_x x\rangle = \langle P_y
y |\mathcal H^{(\rm long)} | P_y y\rangle = \lambda\ ,
\end{equation}
\[
\langle P_x y|\mathcal H^{(\rm long)} | P_x y\rangle = \langle P_y
x
 |\mathcal H^{(\rm long)} | P_y x\rangle =\eta\ ,
\]
\[
\langle P_y x|\mathcal H^{(\rm long)} | P_x y\rangle = \langle P_x
y |\mathcal H^{(\rm long)} | P_y x\rangle = \mu\ ,
\]
where $\lambda = 3\eta = 3\mu$. According to
Eq.~(\ref{psplitting}) and in agreement with the angular-momentum
considerations, the $P$-shell of the bright exciton in an
axisymmetric quantum disk is split into three sublevels, see
Fig.~1. The outermost sublevels labelled $0^U$, $0^L$ are
nondegenerate and characterized by a zero total angular-momentum
$z$-component. The central doubly-degenerate sub\-level
corresponds to the angular-momentum component $\pm 2$. The
intersublevel energy spacing, $\Delta$, equals to $2\eta = 2\mu$.

The splitting $\Delta$ depends on the character of exciton
confinement in a disk. Let us assume $A_{e,h} = \hbar^2/2m_{e,h}
a^4$, where $a$ is the disk effective radius, $m_{e,h}$ are the
electron and hole effective masses. It follows then that in the
strong confinement regime, $a \ll a_{\rm B}$, one has
\begin{equation}\label{delta:sep}
\Delta = \frac{3\sqrt{2\pi}}{16a^3\varkappa_b} \left(\frac{e\hbar
|p_0|}{m_0 E_g} \right)^2.
\end{equation}
In the opposite limiting case $a \gg a_{\rm B}$ where the exciton
is quantized as a whole we obtain
\begin{equation}\label{delta:whole}
\Delta = \frac{3\sqrt{\pi}}{2 a_{\rm exc} a_{\rm B}^2 \varkappa_b}
\left(\frac{e\hbar |p_0|}{m_0 E_g} \right)^2
\end{equation}
with $a_{\rm exc} = a(\mu/M)^{1/4}$ being the radius of exciton
in-plane confinement, $M = m_e + m_h$ and $\mu = m_e m_h/M$.
\section{$P$-Shell excitons in anisotropic quantum disks}
We introduce a slightly elliptical lateral potential replacing the
disk radius $a$ by the effective radii $a_x = a - d$, $a_y = a +
d$ along the $x$ and $y$ axes, respectively, and assuming $d \ll
a$. At zero $d$ the exciton $P$-states are partially split by the
exchange interaction. The anisotropy of a quantum disk results in
a full removal of the degeneracy and formation of four sublevels.
It is convenient to introduce, as a parameter describing the
anisotropy, a half of the $P_x$-$P_y$ splitting, $E_{\rm anis}$,
calculated neglecting the exchange interaction. Its value depends
on the model of exciton quantization. If an electron and a hole
are quantized independently ($a \ll a_{\rm B}$) then, for the
$|P_e, S_h \rangle$ excited state, we have
\begin{equation}\label{omega}
E_{\rm anis} =\frac{d}{a} \frac{2\hbar}{m_e a^2}
\end{equation}
and, for the $|S_e, P_h \rangle$ state, $E_{\rm anis}$ differs
from Eq.~(\ref{omega}) by the replacement $m_e \rightarrow m_h$.
Here $|S_e, P_h \rangle$ means an exciton formed by a $S_e$-shell
electron and a $P_h$-shell hole. If exciton is quantized as a
whole then in Eq.~(\ref{omega}) one should replace $m_e$ by the
exciton translational mass $M = m_e + m_h$ and $a$ by $a_{\rm exc}
= a(\mu/M)^{1/4}$.

Figure 1 shows splitting of the $P$-shell bright-exciton level as
a function the ratio $\xi = E_{\rm anis}/\Delta$. For small values
of $\xi$ the splitting of the doublet $\pm 2$ is proportional to
$\xi^2$. In the limit of strong anisotropy, $E_{\rm anis} \gg
\Delta$, the $P$-shell exciton states form two doublets, $|P_x,
\alpha \rangle$ and $|P_y, \alpha \rangle$ $(\alpha = x,y)$,
separated by $2E_{\rm anis}$ and each split by $\Delta$.

\begin{figure}[ht] \leavevmode\epsfxsize=2.6in
\centering{\epsfbox{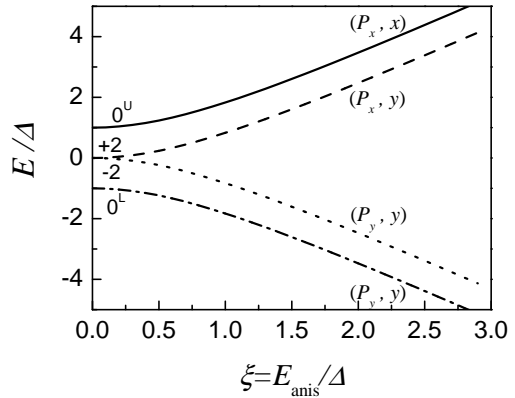}}\label{fig:psplt} \caption{Interplay
between the in-plane anisotropy and exchange interaction. The
exciton sublevel energy $E$ is referred to the energy of the
uniaxial exciton $\pm 2$.}
\end{figure}
\section{Resonant $P$-$S$ excitonic polarons}
Finally, we briefly discuss the $P$-$S$ excitonic polaron which is
formed as a result of the resonant coupling of the $P$- and
$S$-like levels by a longitudinal optical (LO) phonon
\cite{bastard1}. We use an approach developed in
Ref.~\cite{zimmerman} to calculate the LO-assisted resonant
coupling between $S$- and $P$-shell electron states confined in a
quantum dot. For the Fr\"olich interaction, the constant of
exciton-phonon $P$-$S$ coupling is given by
\begin{equation}\label{gamma}
\gamma=\sqrt{\frac{2\pi e^2 \hbar\Omega}{V\varkappa^*}\sum_{\bm q}
\left|\frac{\mathcal I(\bm q)}{q} \right|^2}\ .
\end{equation}
Here $\Omega$ is the LO-phonon frequency, $V$ is the 3D
normalization volume, $\varkappa^{*-1} = \varkappa^{-1}_{\infty} -
\varkappa^{-1}_0$, and
\[
\mathcal{I}(\bm q) = \int d\bm r_e d\bm r_h \Psi_p(\bm r_e, \bm
r_h) \Psi_s(\bm r_e, \bm r_h) \left({\rm e}^{{\rm i} \bm q\bm r_e}
- {\rm e}^{{\rm i} \bm q\bm r_h}\right).
\]

Depolarization of the linearly-polarized $P$-$S$ excitonic polaron
excited via the $P$-shell is determined by the ratio of $\gamma$
and the exchange splitting $\Delta$. If $\Delta \gg |\gamma|$ the
initial linear polarization is being rapidly lost and the exciton
photoluminescence is practically depolarized. On the contrary, if
$\Delta \ll |\gamma|$ then the polarization relaxes on a much
longer time scale and can be well preserved.

The work is supported by RFBR, ``Dynasty'' foundation -- ICFPM,
the Center for Functional Nanostructures of the Deutsche
Forschungsgemeinschaft within project A2 and by the programs of
Russian Academy of Sci.

\end{document}